\documentclass[manuscript]{aastex}
\usepackage{lscape}
\pdfoutput=1

\slugcomment{Accepted for Publication in Planetary and Space Science}

\shorttitle{}
\shortauthors{Kulkarni et al.}

\begin{document}

\title{Probing the Interstellar Dust in Galaxies over $>10$ Gyr of Cosmic History}

\author{Varsha P. Kulkarni}
\affil{University of South Carolina, Dept. of Physics and Astronomy, Columbia, SC 29208}
\email{kulkarni@sc.edu}

\author{Monique C. Aller}
\affil{Georgia Southern University, Dept. of Physics, Statesboro, GA 30460}

\author{Donald G. York and Daniel E. Welty}
\affil{Department of Astronomy \& Astrophysics, University of Chicago, Chicago, IL 60637}

\author{Giovanni Vladilo}
\affil{Osservatorio Astronomico di Trieste}

\author{Debopam Som}
\affil{University of South Carolina, Dept. of Physics and Astronomy, Columbia, SC 29208;  
Aix Marseille Universit\'e, CNRS, Laboratoire dÕAstrophysique de Marseille, UMR 7326, 13388, Marseille, France}

\begin{abstract}
This article is based on an invited talk given by V. P. Kulkarni at the 8th Cosmic Dust meeting. Dust has a profound effect on the physics and chemistry of the interstellar gas in galaxies and on the appearance of galaxies. Understanding the 
cosmic evolution of dust with time is therefore crucial for understanding the evolution of galaxies. Despite the importance of interstellar dust, very little 
is known about its nature and composition in distant galaxies.  We summarize 
the results of our ongoing programs  using observations of distant quasars to obtain better constraints on 
dust grains in foreground galaxies that happen to lie along the quasar sightlines. These observations consist of a combination of mid-infrared data 
obtained with the Spitzer Space Telescope and optical/UV data obtained with ground-based telescopes and/or the Hubble Space Telescope. The 
mid-IR data target the 10 $\mu$m  and 18 $\mu$m silicate absorption features, while the optical/UV data allow determinations of element depletions, extinction curves, 2175 {\AA} bumps, etc. Measurements of such properties in absorption-selected galaxies with redshifts ranging from $z\sim0$ to $z>2$  provide constraints 
on the evolution of interstellar dust over the past $> 10$ Gyr. The optical depth of the 10 $\mu$m silicate absorption feature ($\tau_{10}$) in these galaxies is correlated with the amount of reddening along the sightline. But there are indications [e.g., based on the $\tau_{10}$ /$E(B-V)$ ratio and possible grain crystallinity] that the dust in these distant galaxies differs in structure and composition from the dust in the Milky Way and the Magellanic Clouds. We briefly 
discuss the implications of these results for the evolution of galaxies and their star formation history. 
\end{abstract}

\keywords{}

\section{Introduction}
Only about $1\%$ of the Milky Way interstellar medium (ISM) is in the form of dust, but dust impacts a number of key astrophysical processes, such as ISM heating, cooling,
molecule production and star formation. Moreover, dust grains act as seeds in the growth of planets. Dust also affects derived galaxy properties such as distances and structures, since it attenuates and reddens starlight. Dust even affects dark energy studies, since accurate mapping of the 
cosmic expansion needs corrections for dust extinction. Unfortunately, not much is known about the dust
in distant galaxies, since it is difficult to
observe with direct imaging. As a result, corrections for dust extinction in distant galaxies often assume that the dust is similar to that in the Milky Way or the 
Magellanic Clouds. 

So how can we study dust in distant galaxies? One way to investigate dust in the distant universe is by using the absorption signatures of galaxies against bright
background sources such as quasars or gamma-ray bursts (GRBs). Indeed, afterglows of distant GRBs have been used for observations and modeling of high-redshift dust, including the 2175 {\AA} bump (e.g., Liang \& Li 2009; Zafar et al. 2012). Here we focus on quasars, since they can be observed over longer periods 
than the GRBs, thus allowing follow-up observations with a variety of facilities (e.g., to investigate the carbonaceous and silicate dust as well as the gas 
properties). The quasar light, on its journey 
spanning billions of light-years toward the Earth, passes 
through multiple galaxies and intergalactic regions at distinct redshifts along the sightline. These foreground objects produce absorption lines redshifted by the corresponding amounts in the quasar spectrum. Measurements of these absorption lines can be used to probe the 
objects producing them. This absorption-line technique selects galaxies only by the gas (or dust) cross-section, independent of galaxy brightness or
distance. It thus offers an excellent tool for studying the evolution of gas or dust in galaxies over large fractions of the age of the universe. 

Especially important probes of the interstellar and circumgalactic gas and dust in galaxies are provided by the strongest of the quasar absorption systems, i.e. 
the damped Lyman-$\alpha$ absorbers (DLAs) and the sub-damped Lyman-$\alpha$ absorbers (sub-DLAs). The DLAs are defined as the absorbers 
that have H I 
column density $N_{\rm H I} \ge 2 \times 10^{20}$ cm$^{-2}$. The sub-DLAs have somewhat lower, but still very large, H I column densities 
$10^{19} \le N_{\rm H I} < 2 \times 10^{20}$ cm$^{-2}$. The DLAs and sub-DLAs contain most of the neutral gas available for star formation in the distant universe. The sub-DLAs, despite their lower H~I column densities, appear to be more metal-rich than DLAs (based on the observed gas-phase abundances of typically undepleted or weakly depleted elements, e.g., Kulkarni et al. 2007a, 2010; Som et al. 2015). 
Given the large quantities of H I, and the presence of metals, the DLAs and sub-DLAs are an ideal place to search for interstellar and circumgalactic dust in the distant universe. We have been using multi-wavelength spectroscopic studies of the strong quasar 
absorbers to study interstellar gas and dust in distant galaxies. Our data are obtained with a combination of UV, visible, and IR facilities such as the Hubble Space 
Telescope, the Magellan Clay telescope, the Very Large Telescope, and the Spitzer Space Telescope, and are complemented by data from the Sloan Digital Sky Survey for some objects.

\section{Signatures of Interstellar Dust} To probe the dust in distant galaxies, we focus on three main signatures of dust: 

(1) Element depletions: Comparing the gas-phase abundances of refractory elements 
relative to that of a volatile element, one can
estimate fractions of elements locked up in the solid phase. 

(2) Reddening / extinction of light from background
sources: Comparing the spectra of reddened and unreddened quasars allows the determination of the extinction curve for the foreground galaxy. 

(3) Spectral features expected from dust grains: Observations of specific absorption or emission features expected for dust grains gives a direct 
probe of the dust. The 2175 {\AA} feature is generally associated with carbonaceous dust and the 
10 $\mu$m and 18 $\mu$m features are associated with silicate dust. 

\subsection{Element Depletions} 
The Milky Way ISM shows an overall trend of decreasing gas-phase element abundances with increasing condensation temperature (e.g., Savage \& Sembach 1996; Jenkins 2009). This anti-correlation is generally interpreted as resulting from the elements with higher condensation temperature condensing out first into the solid phase. [We note, however, that some exceptions exist--e.g., some elements (such as Si and Fe) with similar condensation temperatures have different depletions, while some elements (such as O and P) have similar depletion levels, despite showing significantly different condensation temperatures.] In general, cool ISM regions  show stronger depletion levels than warm ISM regions. 

Fig. 1a shows the abundance pattern for a few key elements in the distant galaxies traced by typical DLAs and sub-DLAs (Prochaska \& Wolfe 2002; Meiring et al. 2009), and comparison Milky Way and SMC ISM abundances (e.g., Welty et al. 1997). It is interesting that both DLA and sub-DLA absorbers show depletion of the more refractory elements 
relative to the volatile element Zn. Moreover, the abundance patterns are roughly similar between DLAs and sub-DLAs, and are comparable to the pattern in the Galactic halo ISM. (We note, however, that individual sightlines with large depletions may still exist; such sightlines may be masked in averages across multiple sightlines. Severe depletions in individual velocity components
in a given sightline can also be masked by averaging over the various
components in that sightline.) Fig. 1b shows the abundance of Fe relative to Zn (or S, if Zn is not present) in a more recent sample of DLAs (Kulkarni et al. 2015). A trend of more severe depletion with increasing metallicity is seen (see also Ledoux et al. 2003, Meiring et al. 2006, Vladilo et al. 2011, who reported such a trend as well). A similar trend is also seen in sub-DLAs (e.g., Meiring et al. 2009). These trends can be understood in terms of an increasing dust-to-gas ratio with increasing metallicity. As the total number of metal atoms in an absorbing region increases, more atoms of the refractory elements are available to condense, which leads to stronger depletion of those elements. 

\subsection{Reddening and Extinction}
Interstellar dust displays a range of reddening properties. The ratio $R_{V} = A_{V} / E(B-V)$ varies between $\sim$2 and $\sim$6 in the Milky Way, 
with a typical value of 3.1 (see, e.g., Draine 2003). Larger values of $R_{V}$ imply the presence of larger grains. The ratio of reddening to gas column density  $E(B-V)/N_{H}$ differs in different galaxies. For example,  $E(B-V)/N_{H}$ is observed to be $ \approx 1. 73\times 10^{-22}$ mag cm$^{2}$/H for $R_{V} = 3.1$ in the diffuse interstellar 
regions in the Milky Way. In the Large Magellanic Cloud (LMC) and the Small Magellanic Cloud (SMC), these ratios are about 4.5 $\times 10^{-23}$  mag cm$^{2}$/H and 2.2 $\times 10^{-23}$ mag cm$^{2}$/H, respectively (see, e.g., Draine 2003). More recently, using larger samples, Welty et al. (2012) find $E(B-V)/N_{H} \approx 1. 78\times 10^{-22}$ mag cm$^{2}$/H for the Milky Way, $6.46 \times 10^{-23}$ for the LMC, and $4.37 \times 10^{-23}$ for the SMC. The average $R_{V}$ values for the SMC bar, LMC, and LMC2 supershell regions are observed to be $\approx$ 2.74, 3.41, and 2.76, respectively (Gordon et al. 2003). 

Distant galaxies also redden spectra of background quasars (e.g., Pei, Fall, \& Bechtold 1991; York et al. 2006a). Pei, Fall, \& Bechtold (1991) reported the 
first evidence for dust in galaxies traced by DLAs by comparing the spectra of quasars with foreground DLAs, and those with no foreground DLAs. 
York et al. (2006a) examined 809 Mg II absorbers with $1.0 \le z_{abs} < 1.9$ detected in the quasar spectra from the Sloan Digital Sky Survey (SDSS). They found a clear evidence 
of a small but statistically significant amount of reddening, and showed that the amount of reddening correlates with the metal absorption line 
strengths. This correlation indicates that the reddening is caused by dust in the absorption systems along the sightlines, and 
also shows that there is more dust in the absorbers with higher metal column densities. Fitting the extinction curve of the quasars with the template of an unreddened quasar, they found that most of the cases could be fitted with an 
SMC-like extinction curve with no  2175 {\AA} bump. However, a small fraction of the distant galaxies do show the 2175 {\AA} feature from carbonaceous dust. Khare et al. (2012) confirmed the existence of dust 
in the distant quasar absorber galaxies, using a sample of 1084 absorbers at $2.1 < z < 5.2$ 
with log $N_{\rm H I} > 20.0$ from the SDSS. They found evidence for stronger reddening for the absorbers with stronger gas-phase Si lines 
($E_{B-V} = 0.0097$ for absorbers with $W_{\rm Si\, II \,1526}  \ge 1.0$ {\AA} vs. $E_{B-V} = 0.0039$ for absorbers with $W_{\rm Si\, II \,1526}  < 1.0$ {\AA}). Furthermore, the extinction of the background quasar shows a strong correlation with the dust-phase column density of Fe  in the absorber, and the mean extinction $A_{V}$ per dust-phase Fe atom for the quasar absorbers appears to be quite similar to that found in the Milky Way ISM (Vladilo et al. 2006).

\subsection{Dust Spectral Features}

\noindent {\bf Carbonaceous Absorption:} The 2175 {\AA} bump, believed to be produced by carbonaceous dust, is seen commonly in the Milky Way ISM, although 
with varying degrees of strength. The LMC ISM shows a weaker bump along some sightlines, while most SMC sight lines do not show the bump (but see Lequex et al. 1982, Gordon et al. 2003, Welty et al. 2013 for an example of an unusual SMC sightline that does show the bump and has molecular absorption and diffuse interstellar bands of similar strengths as in the Milky Way). Most quasar absorbers do not show the 2175 {\AA} bump (e.g. York et al. 2006a). The extinction curves for most quasar absorbers appear to be well-fitted by an SMC-like extinction curve. On the other hand, a small fraction of absorbers do show the 2175 {\AA}  bump  (e.g., Junkkarinen et al. 2004; Srianand et al. 2008;
Noterdaeme et al. 2009; Jiang et al. 2010, 2011; Kulkarni et al. 2011; Ledoux et al. 2015; Ma et al. 2015). While most 2175 {\AA} bumps in these studies are relatively weak, similar to the bump toward the LMC, some cases of stronger bumps more resembling the bump in the Milky Way ISM also exist.  Ledoux et al. (2015) report that $\sim 30\%$ of the absorbers with detections of C I absorption show the 2175 {\AA} bump. These absorbers are evidently more rich in carbonaceous dust than most other absorbers. 

\noindent {\bf Silicate Absorption:} Silicates consist of SiO$_{4}$ tetrahedra, combined with Mg$^{++}$ or Fe$^{++}$ ions.  If the silicates are crystalline, the SiO$_{4}$  tetrahedra can share O atoms with other tetrahedra. In amorphous silicates, the number of shared O atoms may be different for 
different tetrahedra. The 10 $\mu$m feature originates in resonances due to Si-O stretching mode in the tetrahedra while the 18 $\mu$m feature originates in resonances due to O-Si-O bending mode. Alignment of the SiO$_{4}$ tetrahedra in crystalline silicates gives rise to sharp peaked resonances, causing sub-structure in the broad 10 $\mu$m and 18 $\mu$m features. On the other hand, amorphous silicates show broad features without substructure (see, e.g., Molster \& Kemper 2005). The silicate features at 10 $\mu$m and 18 $\mu$m are prominent in the ISM of the Milky Way, and are 
observed to be broad and featureless, indicating that the silicate dust is amorphous (see, e.g., Li \& Draine 2001). 

Silicates comprise $\sim 70\%$ mass of the interstellar dust grains in the Milky Way (e.g., Greenberg \& Li 1999). Despite this, the silicate dust was not probed in most past studies of dust in quasar absorber galaxies, which focused primarily on the 
search for the 2175 {\AA} bump. In the past few years, our team made the first detections of silicate dust in absorption-selected galaxies at redshifts up to 
$\sim 1.4$. We now discuss the results from these studies in detail. 

\section{Silicate Dust in Quasar Absorber Galaxies}

We have been carrying out a search for the silicate features in distant galaxies in 
the spectra of background quasars. To do this search, we used the Infrared Spectrograph (IRS) on the Spitzer Space Telescope, since it provided the
necessary wavelength coverage with adequate sensitivity and spectral resolution to search for the redshifted silicate features at 10 $\mu$m and 18 $\mu$m. 

The Spitzer IRS spectra were obtained for quasars that are known to have foreground absorber galaxies at redshifts $0.2 < z < 1.4$. Each of these sightlines is dominated by a single strong absorber. The redshifts of these 
absorbers are known accurately from the measurements of their metal absorption lines in the optical/UV spectra of the background quasars. These specific absorbers were 
selected because they have abundant gas and dust. The specific dust signatures we looked for while selecting these targets were: detection of 2175 {\AA}  absorption, strong reddening or extinction ($A_{V}>0.4$) compared to typical quasar absorbers, or strong element depletions. Some of our targeted absorbers also show detection of molecules (which is, again, highly unusual compared to most other quasar absorbers, since only a very small fraction of absorbers are known to have detections of molecules). The IRS spectra covered the 10 $\mu$m feature in every galaxy studied and a part of the 18 $\mu$m feature in 6 galaxies. 

\subsection{Detections of the 10 $\mu$m Feature} 
The first detection of silicate absorption in a quasar absorber was made by Kulkarni et al. (2007). They observed a $z_{abs}=0.52$ absorber along the sightline to the $z_{em}=0.94$ blazar AO 0235+164 (Fig. 2).  This DLA absorber offers an excellent place to search for silicate dust because 
it has one of the highest H I column densities (log $N_{\rm H I} = 21.7$), shows 21-cm absorption, X-ray absorption, a 2175 {\AA} bump, and detection of 
diffuse interstellar bands (Roberts et al. 1976, Junkkarinen et al. 2004, York et al. 2006b). The background blazar also shows substantial reddening 
[$E(B-V) = 0.23$ in the absorber rest frame]. The Spitzer IRS spectrum showed a broad, shallow 10 $\mu$m silicate feature at a $\sim$15 $\sigma$ level. Surprisingly, the peak optical depth of the feature ($\tau_{10} \approx 0.08$) is about twice that expected based on an extrapolation of the $\tau_{10}$ vs. $E(B-V)$ relation for the diffuse interstellar clouds in the Milky Way. The profile of the 10 $\mu$m feature is fitted reasonably well by a template corresponding to 
the profile for a diffuse Galactic ISM cloud or the profile for laboratory amorphous olivine. 

This initial detection was followed up by Spitzer IRS observations of four more quasar absorbers. The 10 $\mu$m silicate absorption feature was detected in all systems  at $> 4 \sigma$ level (Kulkarni et al. 2011). Several more absorbers have been observed since then, and show the 10 $\mu$m absorption (Aller et al. 2012, 2013, 2014, 2016). In general, these absorbers also show the peak optical depth of the 10 $\mu$m feature to be 
2-3 times larger than expected based on an extrapolation of the $\tau_{10}$ vs. $E(B-V)$ relation for the diffuse interstellar clouds in the Milky Way. 
Thus the dust in these distant galaxies is likely to be more silicate-rich than that in the diffuse clouds of the Milky Way ISM. 

\subsection{Crystalline Silicates} 
The sightline toward the blazar PKS 1830-211 is interesting because of the presence of an intervening face-on spiral galaxy at redshift $z_{abs} = 0.89$ with a strong molecular content. A number of molecules, such as CO,  CS, H$_{2}$O, NH$_{3}$, HCN, H$^{13}$CN, CH$_{3}$CHO, and CH$_{3}$NH$_{2}$, have been detected in absorption in this galaxy (e.g., Wiklind \& Combes 1996, 1998; Menten et al. 2008; Muller et al. 2011 and references therein). This absorber was therefore targeted for our Spitzer IRS search for silicate dust. Strong absorption is indeed detected in the 10 $\mu$m feature (Aller et al. 2012). However, this feature is not fit well by the smooth template for amorphous olivine, since it exhibits more structure. We attempted to fit the observed 10 $\mu$m profile with more than 100 templates from laboratory measurements and Galactic as well as extragalactic astrophysical sources. The best-fit to the observed profile is provided by a laboratory crystalline olivine (Hortonolite, i.e. Mg$_{1.1}$Fe$_{0.9}$SiO$_{4}$; Fig. 3). This source appears to show $\ge 95\%$ crystallinity, even higher than the highest crystalline mass fraction of $\sim 90\%$ ever reported in silicate-producing evolved stars (Jiang et al. 2013). 

Such high crystallinity in the interstellar/circumgalactic gas in the absorber galaxies is very surprising. Highly crystalline silicate dust has been detected in comets ($\sim$7-90$\%$ crystallinity; e.g. Wooden et al. 1999; Bouwman et al. 2003; Molster \& Kemper 2005) and post-AGB stars ($\sim$10-75$\%$ crystallinity; e.g., Molster et al. 2001; Molster \& Kemper 2005). 
However, the Milky Way ISM shows mostly a broad 10 $\mu$m feature 
indicative of amorphous silicates. The crystallinity of the 
diffuse ISM is estimated to be $\le 2-5\%$ (Li \& Draine 2001; Kemper et al. 2004; Li et al. 2007). Bowey \& Adamson (2002) examined whether the broad 10 $\mu$m feature could result from a 
superposition of a large number of crystalline silicate features. However, this appears to be unlikely, given that 
the crystalline silicate resonance features at longer wavelengths are not observed (Molster \& Kemper 2005). 

Crystalline silicate dust emission has been reported in active galaxies (e.g., Xie et al. 2014). Some crystallinity has also been reported in ultra-luminous infrared galaxies (ULIRGs, Spoon et al. 2006), but this is only of the order of 6-13$\%$. 
Furthermore, the crystalline features in ULIRGs are more noticeable at wavelengths longward of the $ 10 $ $\mu$m feature. Spoon et al. attributed the 6-13$\%$
crystallinity (which is still higher than that seen in the Milky Way ISM) to a short timescale for the injection of crystalline silicates into the ISM in the merger-driven starbursts in the ULIRGs. But Kemper et al. (2011) suggest that the crystallinity in these sources may be lower, or that it may be partly associated with the active galactic nucleus. In any case, while at least some connection between silicate crystallinity and the starburst activity seems plausible in ULIRGS, there is no sign of strong star formation activity in the host galaxy of the absorber toward PKS1830-211. 
Thus, the absorber toward PKS1830-211 appears to be unusually rich in crystalline silicates. 
Examination of the 10 $\mu$m features in the other quasar absorbers analyzed so far suggests that some of them may also have crystalline 
silicates, although at a lower level, typically $\sim 30 \%$ (Aller et al. 2013, 2015). 

\subsection{The 18 $\mu$m Feature}

The 18 $\mu$m silicate feature arises in O-Si-O bending modes. The ratio of the strengths of the 10 and 18 $\mu$m features varies with changes in dust grain properties such as the distribution of grain shapes and sizes. 
In some of the quasar sightlines, the IRS data covered the 18 $\mu$m feature as well as the 10 $\mu$m feature. Fig. 4 shows the 10 and 18 $\mu$m features in the absorber at $z=0.68$ toward TXS 0218+357 (Aller et al. 2014). The ratio of the peak optical depths of these features in this sightline is $\tau_{10} / \tau_{18} = 1.31 \pm 0.48$. The best-fitting template for this object is that of amorphous olivine with porous,  ellipsoidal particles (Chiar \& Tielens 2006, although we note that given the higher noise level in this spectrum, we cannot rule out the existence of substructure expected for crystalline silicates). The $\tau_{10} / \tau_{18}$ ratio for this best-fitting template is 1.41, which matches the observed value better than the ratio expected for an amorphous olivine template with solid, spherical particles (which predicts $\tau_{10} / \tau_{18} = 1.62$). Future observations covering wavelengths longward of the 18 $\mu$m feature are essential to obtain more accurate measurements of the 18 $\mu$m feature and hence of the $\tau_{10} / \tau_{18}$ ratio. It is interesting to note, however, that a wide range of values are observed for this ratio. For example, silicate absorption features in 
sightlines toward WR stars with high extinction values show $\tau_{10} / \tau_{18}$ $\sim  1.4-2.0$ (e.g., Chiar \& Tielens 2006). On the other hand, Hao et al. (2005), who detected 10 and 18 $\mu$m emission features in quasars, measured the ratio $\tau_{10} / \tau_{18}$ to be in the range 0.30 to 0.93. 

\subsection{Silicate Strength Correlations with Other Absorber Properties}

In the Milky Way ISM, the strength of the silicate absorption is well-correlated with the reddening of background stars. In the diffuse ISM of the Milky Way, this trend is well-fit by the relation $\tau_{10} = 0.17 E(B-V)$ (Roche \& Aitken 1984). Extrapolating this trend to the low reddening 
values for the quasar absorbers, one can estimate the expected $\tau_{10}$ values in these sightlines. These estimates are much lower than 
the observed $\tau_{10}$ values for these distant galaxies, as mentioned earlier (Kulkarni et al. 2011, Aller et al. 2014, 2015). This suggests that 
these distant galaxies are more rich in silicate dust, possibly indicating more O-rich environments. It is also possible that these sightlines probe inner regions of the foreground galaxies, given that the silicate feature in the central 
region of the Milky Way is also known to be stronger than in the diffuse ISM in the outer parts of the Milky Way. Roche \& Aitken (1985) suggested that the latter difference may be a result of the Galactic Center region having fewer carbon stars. On the other hand, the 3.4 $\mu$m feature (originating in the C--H stretch from aliphatic hydrocarbon dust) is also stronger in the Galatic Center region, with the $\tau_{3.4}$/$A_{V}$  
ratio being about twice as high toward the Galactic Center as that in the local ISM (e.g., Pendleton et al. 1994, Sandford et al. 1995). Gao et al. (2010) suggest that this difference between the Galactic Center and the local ISM arises because the dense molecular clouds toward the Galactic Center contain porous composite dust. Whatever the underlying reason for the difference between the Galactic Center and the local ISM sightlines, it is possible that the sightlines covered in the studies of silicate absorption in quasar absorbers so far probe regions that resemble the Galactic Center more than the local ISM. Another possibility is that the grains in the quasar absorbers may be larger in size resulting in  lower UV extinction. 

The silicate absorption strength appears to be correlated with the gas-phase metal line strength. For example, $\tau_{10}$ seems to be generally correlated with the rest-frame equivalent width of the Mg II $\lambda$2796 absorption line. This Mg II line is commonly observed in quasar absorbers and is often highly saturated. Its equivalent width is thus more indicative of the velocity dispersion of the absorbing gas rather than of the Mg II column density. A higher velocity dispersion could reflect a higher mass for the absorbing galaxy. On the other hand, recent studies indicate that strong Mg II absorbers may be associated with gas outflows in star-forming galaxies 
(e.g., Bouch\'e et al. 2012, Kacprzak et al. 2012, Bordoloi et al. 2014). This suggests that the absorbers with stronger silicate 10 $\mu$m absorption may be associated with more massive galaxies or with more vigorously star-forming galaxies exhibiting stronger outflows. While the absorbing galaxy toward PKS1830-211 does not display strong star formation, it is essential to measure the star formation rates (SFRs) in the host galaxies of more silicate absorbers to establish whether or not the silicate absorption strength correlates with the SFR.

There is a very tentative suggestion that the silicate dust absorption strength may be anti-correlated with the carbonaceous dust absorption
strength (Kulkarni et al. 2011). Such an anti-correlation, if confirmed, may indicate the dichotomy between C-rich environments and O-rich environments for dust formation. Finally, there is also a tentative anti-correlation between the silicate peak optical depth and the 21-cm optical depth (Aller et al. 2014). Larger samples are needed to verify more definitively whether these trends really exist. 

\section{Conclusions and Discussion}

Silicate dust appears to be common in Spitzer IRS observations of distant galaxies. The silicate absorption is seen at 10 $\mu$m (and 18 $\mu$m) in every target where these features are covered. 
There appears to be considerable variation in the shape, center, and breadth of the silicate absorption features (see Fig. 5). Moreover, the data suggest the presence of crystalline silicate grains in some quasar absorbers. The silicate absorption strength appears to be correlated with reddening, with a steeper slope than in Milky Way ISM. A correlation is also seen between the silicate dust absorption strength and the Mg II gas absorption strength, suggesting that silicate-rich galaxies may be more massive or may have stronger outflows. A correlation between 
stronger outflows and stronger silicate absorption may seem surprising if outflows destroy dust grains. On the other hand, 
stronger outflows may be associated with higher SFRs and more massive stars resulting in more type II supernovae, leading to higher silicate dust production. 

The presence of crystalline grains in some of these distant galaxies, if
confirmed, would show striking differences from the dust in the
Milky Way ISM, which has amorphous silicate dust. Amorphous grains result from condensation at temperatures $T <
1000$ K followed by rapid cooling. Crystalline grains result from condensation at higher
temperatures followed by slow cooling. After injection into the ISM, crystalline grains can change into
amorphous grains because of exposure to cosmic rays. If distant galaxies have more crystalline dust, perhaps
it arises in warmer regions with lower cosmic ray flux (due to lower star formation rates and supernova rates).  

It thus appears that the dust in distant galaxies may differ in significant ways from the dust in the Milky Way and the Magellanic Clouds. 
To further establish the properties of the dust grains in distant galaxies, it is essential to expand the samples and also to explore the connections between the dust properties and the gas properties. For example, it would be very useful to determine the depletions of Si, Fe, and Mg in absorbers with detections of silicate dust. 
We are currently working on the dust-gas connections using data archives from Spitzer, Herschel and ground-based telescopes. The archival data will 
also be used to measure shapes of extinction curves in individual quasar absorbers and to constrain relative abundances of silicate and carbonaceous dust grains. The overall aim of this ongoing archival study [being carried out in collaboration with Dr. E. Dwek (NASA GSFC)] is to constrain the cosmic evolution of carbonaceous and silicate dust. It would also be interesting to test potential connections between higher silicate crystallinity and lower cosmic ray flux with measurements of the host galaxy SFRs and the gas kinematics for the absorbers with crystalline silicate dust.

Several open questions still remain in the study of interstellar dust  in both, the nearby universe and the distant universe: e.g., where and how grains are formed, how they are transported from
formation sites to interstellar clouds, how they are processed (radiation, shocks)
along the way, how they are built up in denser clouds, and what their structure is. To make progress on resolving these broad questions, it 
will help to tackle specific issues such as the deviations from the mean depletion vs. condensation temperature trend, the evolution of carbonaceous vs. silicate dust, the influence of dust formation vs. dust processing on grain crystallinity, and the differences between the dust in quasar DLAs and GRB DLAs (which are believed to represent environments differing in star formation rates and in distances from the galaxy centers). Ongoing and future multi-wavelength observations of quasars and GRB afterglows promise to shed more light on these issues. Characterizing the differences between the dust grains in these distant galaxies and those in the Milky Way and nearby galaxies will be vital to better understand the chemical evolution of galaxies and to more accurately measure the rate of cosmic expansion at earlier epochs. 

\acknowledgments
We thank two anonymous referees for comments that have helped to improve this paper. VPK thanks the organizers of the Cosmic Dust VIII meeting for the invitation and for organizing this wonderful and stimulating 
meeting. VPK, MCA, and DS acknowledge partial support from the National Science Foundation grant AST/1108830, NASA   grant  
NNX14AG74G, and NASA grants issued by JPL/Caltech in support of our Spitzer General Observer programs. Additional support from NASA Space Telescope Science Institute grant HST-GO-12536.01-A and NASA Herschel Science Center grant 1427151 is gratefully acknowledged. DS also acknowledges support from the A*MIDEX project (ANR- 11-IDEX-0001-02) funded by the 
``Investissements d'Avenir" French Government program, 
managed by the French National Research Agency (ANR).

\clearpage

\begin{figure}
\epsscale{1.10}
\plottwo{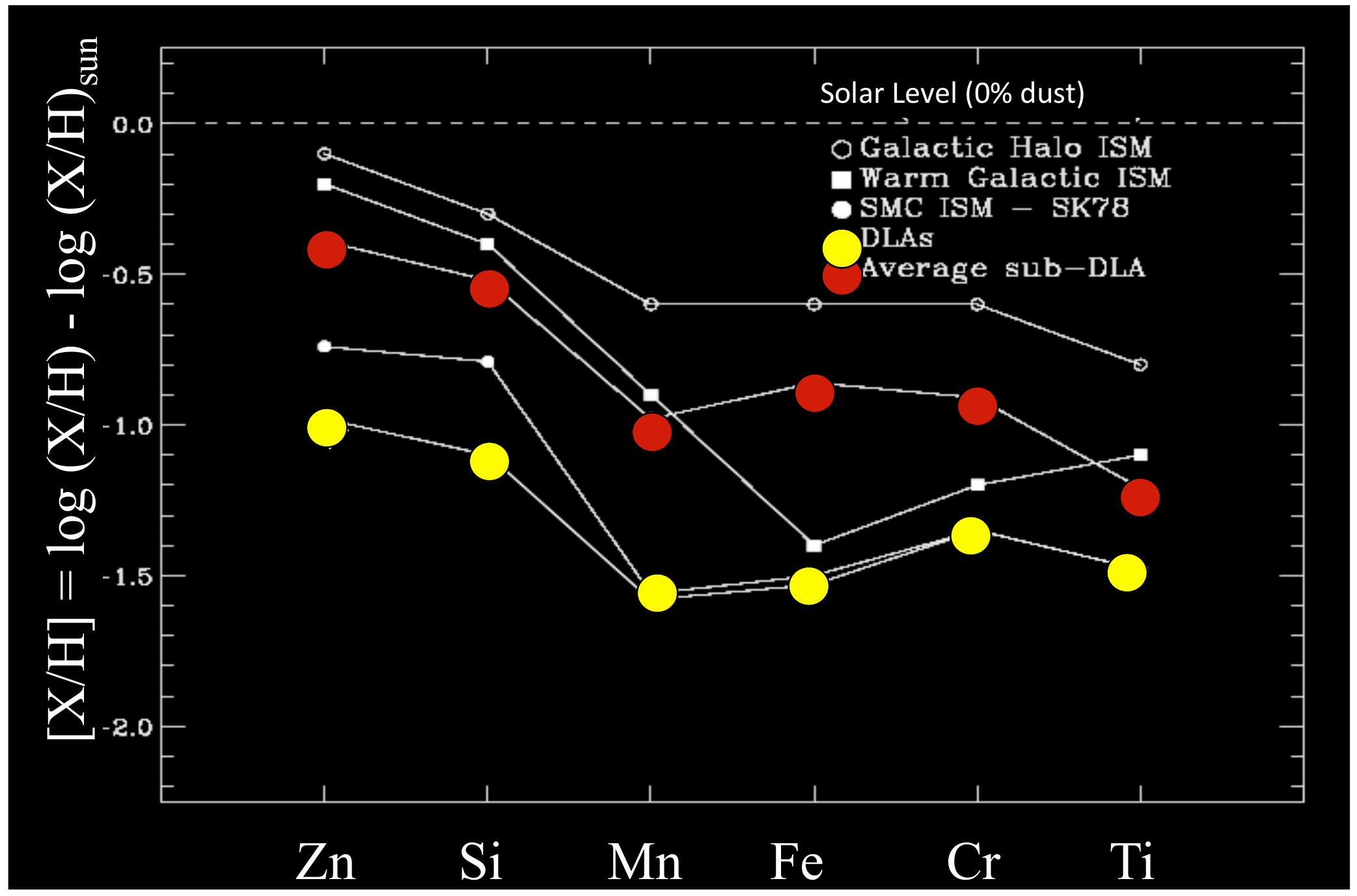}{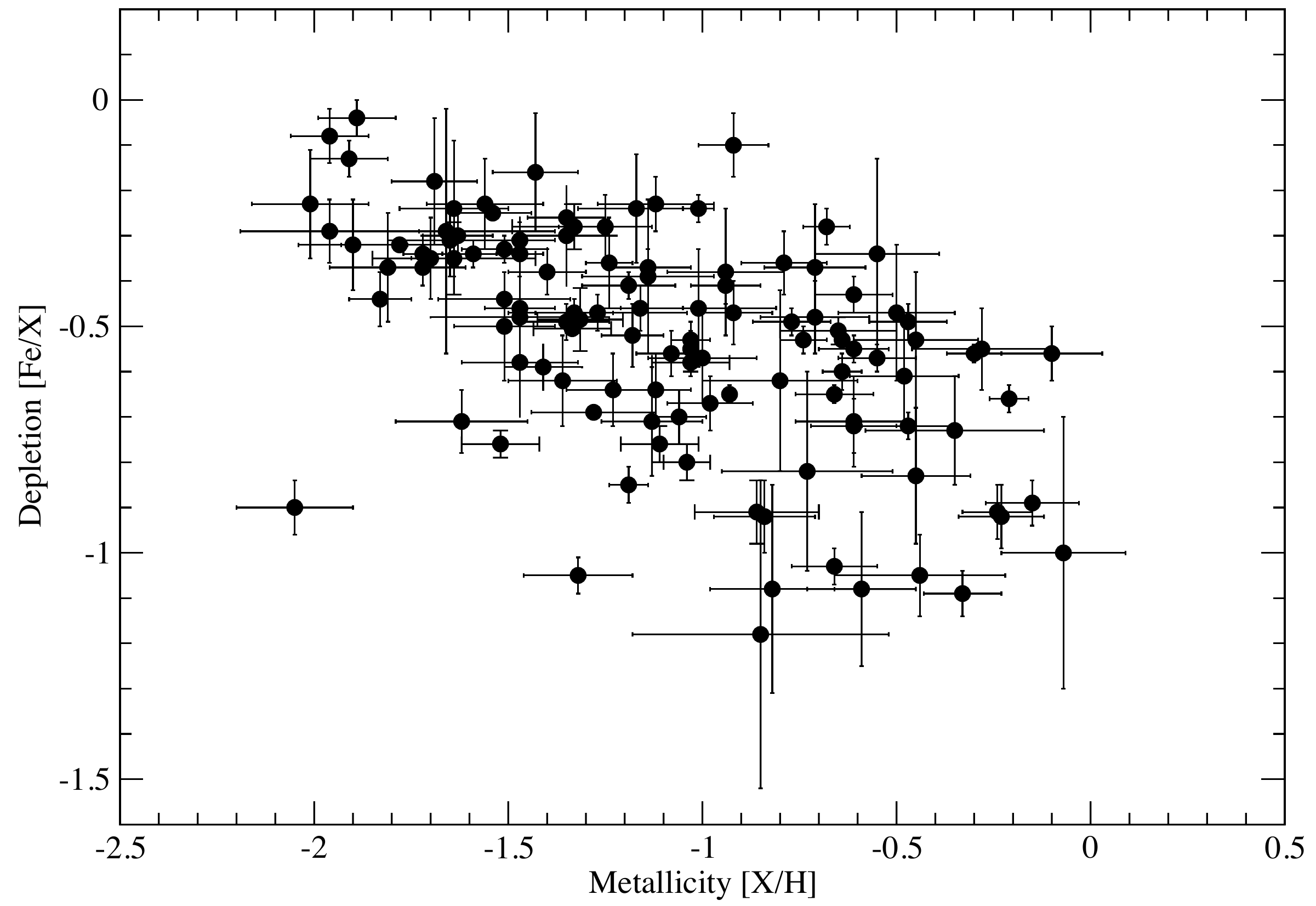}
\caption{Left panel: (a) Abundances of a few key elements in DLAs and sub-DLAs, based on data from Meiring et al. (2009) and references therein. Also shown for comparison are the abundance patterns observed for the Galactic halo ISM, warm Galactic ISM, and the SMC ISM. The depletion patterns are similar for DLAs and sub-DLAs, except the pattern for sub-DLAs is offset to higher abundances. Both the DLA and sub-DLA depletion patterns appear to be roughly similar to the Galactic halo ISM pattern. Right panel: (b)  Depletion of Fe relative to Zn or S (if Zn is not available) vs. the metallicity based on Zn or S, for the sample of quasar DLAs analyzed in Kulkarni et al. (2015). A trend of increasing depletion level with increasing metallicity is seen. \label{fig1}}
\end{figure}

\begin{figure}
\epsscale{0.70}
\plotone{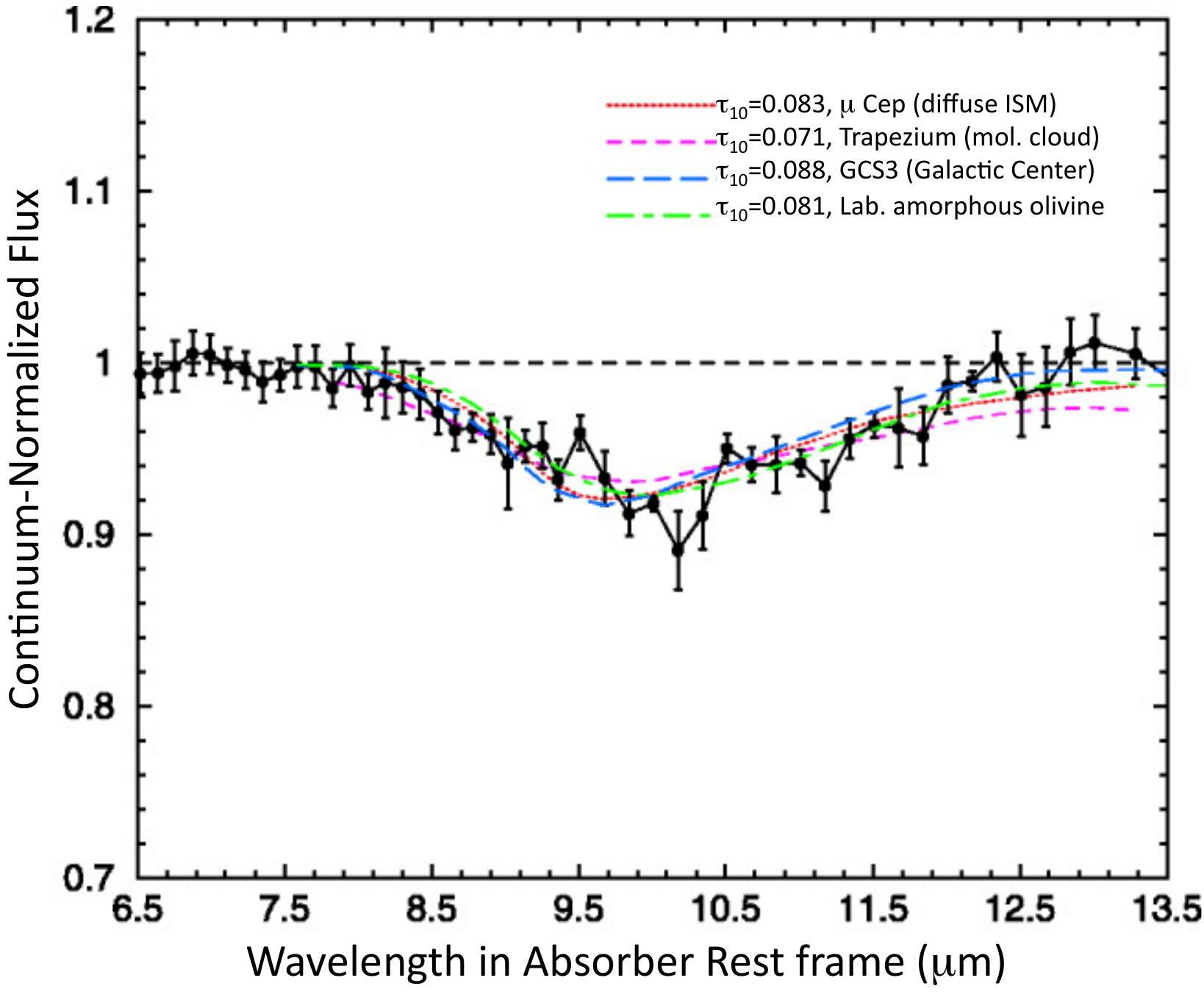}
\caption{ The first detection of silicate 10 $\mu$m absorption in an absorption-selected galaxy at $z=0.52$ toward AO 0235+164 (based on data from Kulkarni et al. 2007b). The curves show the astrophysical and laboratory templates fitted, along with the peak optical depth obtained for teach template. Laboratory amorphous olivine provides a reasonable fit for this absorber. \label{fig2}}
\end{figure}

\begin{figure}
\epsscale{0.70}
\plotone{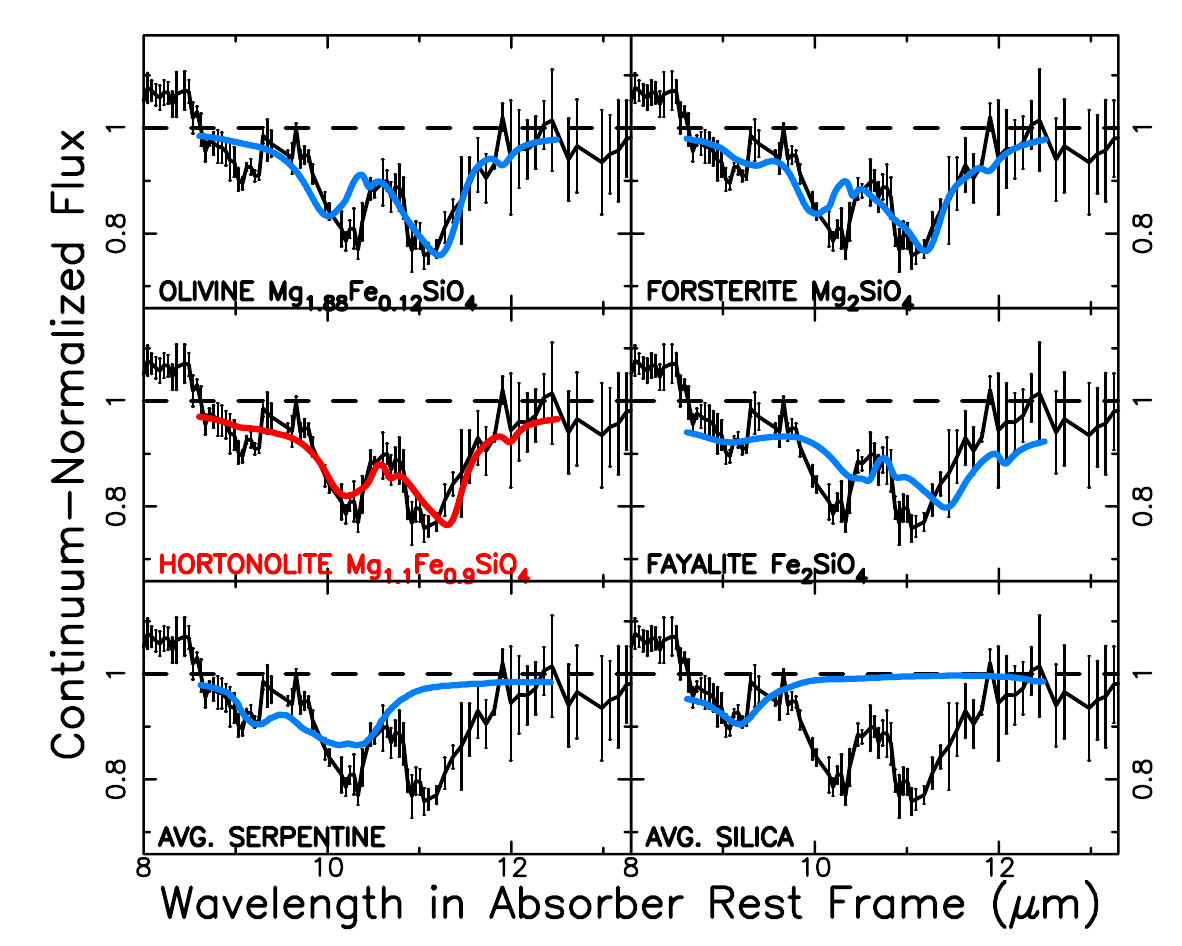}
\caption{ The silicate 10 $\mu$m absorption feature in a quasar absorber at $z=0.89$ toward PKS 1830-211 (based on data from Aller et al. 2012). This absorber 
appears to show crystalline silicates. Fits to the observed profile with a few of the template spectra are shown. The best-fitting template is that for Hortonolite (Mg$_{1.1}$Fe$_{0.9}$SiO$_{4}$). \label{fig3}}
\end{figure}

\begin{figure}
\epsscale{1.00}
\plotone{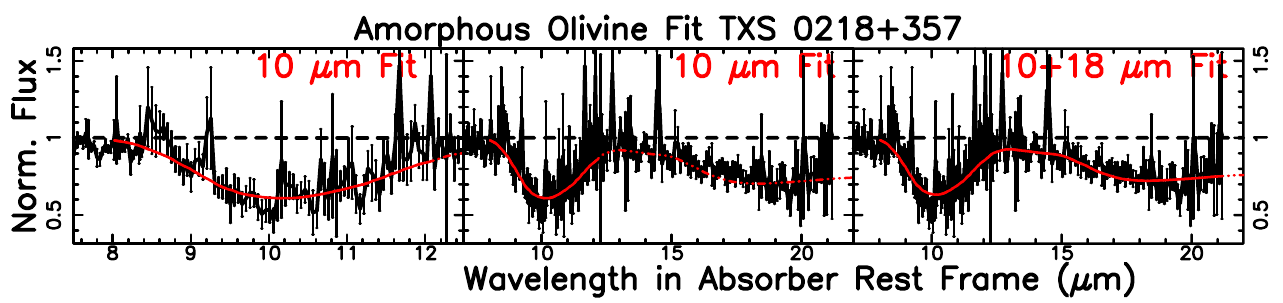}
\caption{ The 10 and 18 $\mu$m absorption features at $z=0.68$ toward TXS 0218+357 (based on data from Aller et al. 2014). Red curves show the best-fitting laboratory amorphous olivine template profiles. The left and center panels show the fits based on the 10 $\mu$m feature only, while the right panel shows the result obtained from simultaneously fitting both the 10 and 18 $\mu$m features. There is no strong evidence for crystalline silicates in this absorber. The ratio of the two features is measured to be $\tau_{10} / \tau_{18} = 1.31 \pm 0.48$. \label{fig4}}
\end{figure}

\begin{figure}
\epsscale{0.9}
\plotone{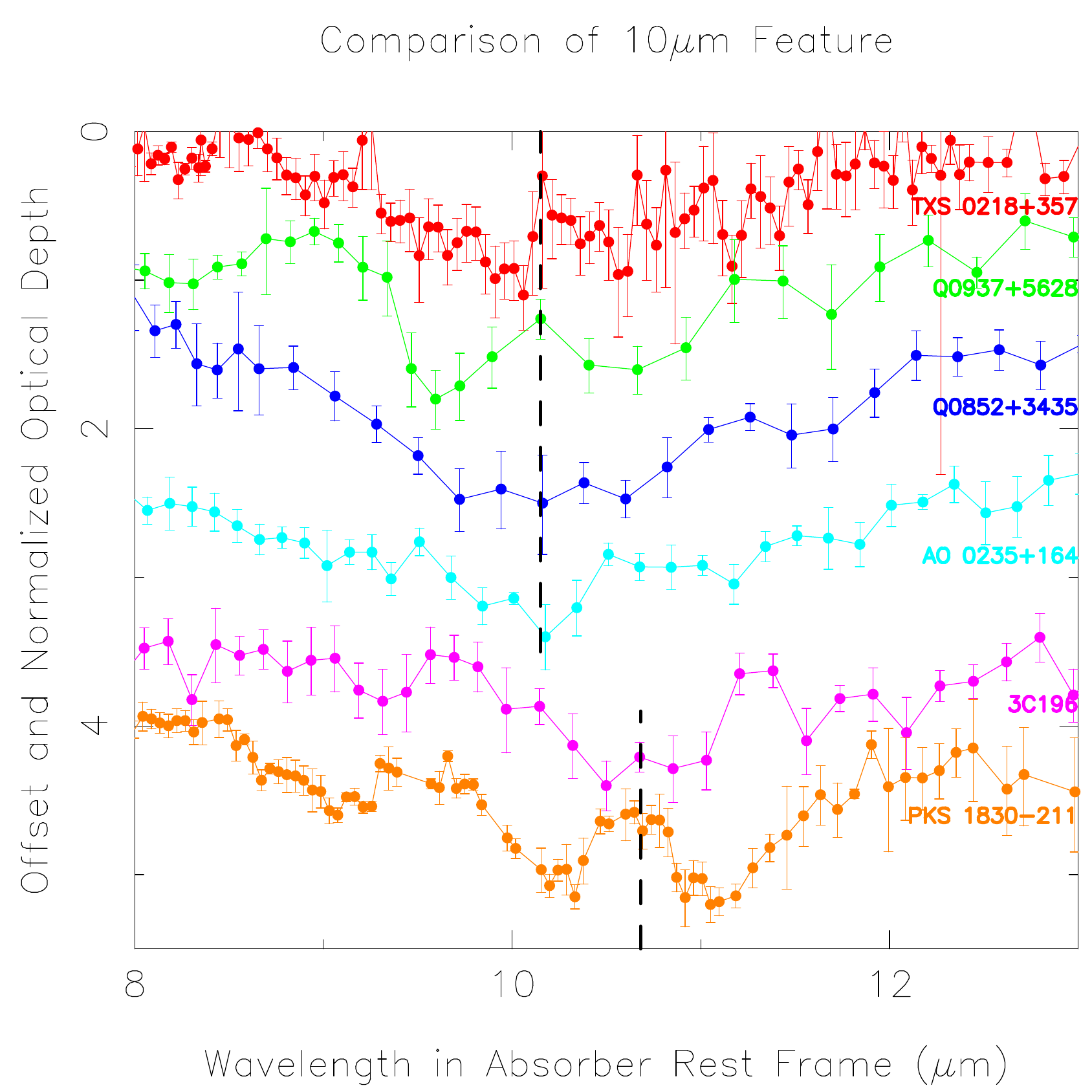}
\caption{ A comparison of the 10 $\mu$m absorption features in some quasar absorbers. There is considerable variation the shape, center, and width of the feature, indicating differences in dust grain properties such as grain shape and composition. \label{fig5}}
\end{figure}
\clearpage					

\begin{thebibliography}{}
\bibitem[Aller12]{Aller12} Aller, M. C., Kulkarni, V. P., York, D. G., Vladilo, G., Welty, D. E., \& Som, D. 2012, ApJ, 748, 19
\bibitem[Aller13]{Aller13} Aller, M. C., Kulkarni, V. P., York, D. G., Welty, D. E., Vladilo, G., \& Som, D. 2013, in Proc. of Science,``The Life Cycle of Dust in the Universe: Observations, Theory, and Laboratory Experiments'', Taipei, Taiwan, Eds. Anja Andersen, Maarten Baes, Haley Gomez, Ciska Kemper, and Darach Watson, id. 13 (arXiv: 1405.0426)
\bibitem[Aller14]{Aller14} Aller, M. C., Kulkarni, V. P., York, D. G., Welty, D. E., Vladilo, G., \& Liger, N. 2014, ApJ, 785, 36
\bibitem[Aller16]{Aller15} Aller, M. C., Kulkarni, V. P.,  York, D. G., Welty, D. E., \& Vladilo, G. 2016, to be submitted
\bibitem[Boi98]{Boi98} Boiss\'e, P., Le Brun, V., Bergeron, J., \& Deharveng, J.-M. 1998, A\&A, 333, 841
\bibitem[Bordoloi14]{Bordoloi14} Bordoloi, R., Lilly, S. J., Kacprzak, G. G., \& Churchill, C. W. 2014, ApJ, 784, 108
\bibitem{Bouche12} Bouch\'e, N., Hohensee, W., Vargas, R., Kacprzak, G. G., Martin, C. L., Cooke, J., \& Churchill, C. W. 2012, MNRAS, 426, 801
 \bibitem[Bouwman03]{Bouwman03} Bouwman, J., de Koter, A., Dominik, C., \& Waters, L. B. F. M. 2003, A\&A, 401, 577 
\bibitem[Bowey02]{Bowey02} Bowey, J. E., \& Adamson, A. J. 2002, MNRAS, 334, 94
\bibitem[Chiar06]{Chiar06} Chiar, J. E., \& Tielens, A. G. G. M. 2006, ApJ, 637, 774
\bibitem[Draine03]{Draine03} Draine, B. T. 2003, ARAA, 41, 241
\bibitem[Gao10]{Gao10} Gao, J., Jiang, B. W., \& Li, A. 2010, Earth Planets Space, 62, 63
\bibitem[Greenberg]{Greenberg} Greenberg, J. M. \& Li, A. 1999, Ad. Space Res., 24, 497 
\bibitem{Hao05} Hao, L., Spoon, H. W. W., Sloan, G. C., Marshall, J. A., Armus, L., Tielens, A. G. G. M., Sargent, B., van Bemmel, I. M., Charmandaris, V., Weedman, D. W., \& Houck, J. R.  2005, ApJL, 625, L75
\bibitem[Jenkins09]{Jenkins09} Jenkins, E. B. 2009, ApJ, 700, 1299
\bibitem[Jiang10]{Jiang10} Jiang P., Ge J., Prochaska J. X. et al. 2010, ApJ, 720, 328
 \bibitem[Jiang11]{Jiang11} Jiang P., Ge J., Zhou H., Wang J., \& Wang T., 2011, ApJ, 732, 110
\bibitem[Jiang13]{Jiang13} Jiang, B. W., Zhang, K., Li, A., \& Lisse, C. M. 2013, ApJ, 765, 72
\bibitem[Kac12]{Kac12} Kacprzak, G. G., Churchill, C. W., \& Nielsen, N. M. 2012, ApJ, 760, L7 
\bibitem[Kemper04]{Kemper04} Kemper, F., Vriend, W. J., \& Tielens, A. G. G. M. 2004, ApJ, 609, 826
\bibitem[Kemper11]{Kemper11} Kemper, F., Markwick, A. J., \& Woods, P. M. 2011, MNRAS, 413, 1192
\bibitem[Kennicutt(1998)]{Ken98} Kennicutt, R. C. 1998,  ApJ, 498, 541
\bibitem[Khare12]{Khare12} Khare, P., vanden Berk, D., York, D. G., Lundgren, B., \& Kulkarni,V. P. 2012, MNRAS, 419, 1028
\bibitem[Kulkarni07a]{Kulkarni07a}  Kulkarni, V. P., Khare, P., P\'eroux, C., York, D. G., Lauroesch, J. T., \& Meiring, J. D. 2007a, ApJ, 661, 88
\bibitem[Kulkarni07b]{Kulkarni07b} Kulkarni, V. P., York, D. G., Vladilo, G., \& Welty, D. E. 2007b, ApJL, 663, L81
\bibitem[Kulkarni10]{Kulkarni10} Kulkarni, V. P., Khare, P., Som, D., Meiring, J., York, D. G., P\'eroux, C., \& Lauroesch, J. T. 2010, NewA, 15, 735
\bibitem[Kulkarni11]{Kulkarni11} Kulkarni, V. P., Torres-Garcia, L. M., Som, D., York, D. G., Welty, D. E., \& Vladilo, G. 2011, ApJ, 726, 14
\bibitem[Kulkarni2012]{KULKARNI12} Kulkarni, V. P., Meiring, J., Som, D., P\'eroux, C., York, D. G., Khare, P., \& Lauroesch, J. T. 2012, ApJ, 749, 176
\bibitem[Kulkarni15]{Kulkarni15} Kulkarni, V. P., Som, D., Morrison, S., P\'eroux, C.,  Quiret, S., \& York, D. G. 2015, ApJ, 815, 24
\bibitem[Ledoux03]{Ledoux03} Ledoux, C., Petitjean, P., \& Srianand, R. 2003, MNRAS, 346, 209
\bibitem[Ledoux15]{Ledoux15} Ledoux, C., Noterdaeme, P., Petitjean, P., \& Srianand, R. 2015, A\&A, 580, A8
\bibitem[Li01]{Li01} Li, A., \& Draine, B. T. 2001, ApJL, 550, L213
\bibitem[Li07]{Li07} Li, M. P., Zhao, G., \& Li, A. 2007, MNRAS, 382, L26
\bibitem[Liang09]{Liang09} Liang, \& Li, A. 2009, ApJ, 690, L56
\bibitem[Ma15]{Ma15} Ma, J., Caucal, P., Noterdaeme, P., Ge, J., Prochaska, J. X., Ji, T., Zhang, S., Rahmani, H., Jiang, P., Schneider, D. P., Lundgren, B., \& Paris, I. 2015, MNRAS, 454, 1751
\bibitem[Meiring06]{Meiring06} Meiring, J. D., Kulkarni, V. P., Khare, P., Bechtold, J., York, D. G., Cui, J., Lauroesch, J. T., Crotts, A. P. S., \& Nakamura, O. 2006, MNRAS, 370, 43 
\bibitem[Meiring09]{Meiring09} Meiring, J. D., Lauroesch, J. T., Kulkarni, V. P., P\'eroux, C., Khare, P., \& York, D. G. 2009, MNRAS, 397, 2037 
\bibitem[Menten08]{Menten08} Menten, K. M., Gusten, R., Leurini, S., Thorwirth, S., Henkel, C., Klein, B., Carilli, C. L., \& Reid, M. J. 2008, A\&A, 492, 725
\bibitem[Moster01]{Molster01} Molster  F. J., Yamamura, I., Waters, L. B. F., Nyman, L.-A. Kaufl, H.-U., de Jong, T., \& Loup, C. 2001, A\&A, 366, 923
\bibitem[Molster05]{Molster05} Molster, F., \& Kemper, C. 2005, Space Sci. Rev., 119, 3
\bibitem[Muller11]{Muller11} Muller, S., Beelen, A., Gu\'elin, M., Aalto, S., Black, J. H., Combes, F., Curran, S. J., Theule, P., \& Longmore, S. N.  2011, A\&A, 535, A103
\bibitem[Pei91]{Pei91} Pei, Y. C., Fall, S. M., \& Bechtold, J. 1991, ApJ, 378, 6
\bibitem[Pendleton94]{Pendleton94} Pendleton, Y. J., Sandford, S. A., Allamandola, L. J., Tielens, A. G. G. M., \&  Sellgren, K. 1994, ApJ,  437, 683
\bibitem[Pettini97]{Pettini97} Pettini, M., Smith, L. J., King, D. L., \& Hunstead, R. W. 1997, ApJ, 486, 665
\bibitem[PW02]{PW02} Prochaska, J. X., \& Wolfe, A. M. 2002, ApJ, 566, 68
\bibitem[Roche84]{Roche84} Roche, P. F., \& Aitken, D. K. 1984, MNRAS, 208, 481
\bibitem[Roche85]{Roche85} Roche, P. F., \& Aitken, D. K. 1985, MNRAS, 215, 425
\bibitem[Sandford95]{Sandford95} Sandford, S. A., Pendleton, Y. J., \& Allamandola, L. J. 1995, ApJ, 440, 697
\bibitem[Som15]{Som15} Som, D., Kulkarni, V. P., Meiring, J.,  York, D. G., P\'eroux, C., Lauroesch, J. T., Aller, M. C., \& Khare, P.2015, ApJ, 806, 25
\bibitem[Spoon06]{Spoon06} Spoon, H. W. W., Tielens, A. G. G. M., Armus, L., Sloan, G. C., Sargent, B., Cami, J., Charmandaris, V., Houck, J. R., 
\& Soifer, B. T.  2006, ApJ, 638, 759
\bibitem[Vladilo06]{Vladilo06} Vladilo, G., Centurion, M., Levshakov, S. A., P\'eroux, C., Khare, P., Kulkarni, V. P., \& York, D. G. 2006, A\&A, 454, 151
\bibitem[Vladilo11]{Vladilo11} Vladilo, G., Abate, C., Yin, J., Cescutti, G., \& Matteucci, F. 2011, A\&A, 530, A33
\bibitem[Welty97]{Welty97} Welty, D. E., Lauroesch, J. T., Blades, J. C., Hobbs, L. M., \& York, D. G. 1997, ApJ, 489, 672
\bibitem[Welty12]{Welty12} Welty, D. E., Xue, R., \& Wong, T.  2012, ApJ, 745, 173
\bibitem[Welty13]{Welty13} Welty, D. E., Howk, J. C., Lehner, N., \& Black, J. H. 2013, MNRAS, 428, 1107
\bibitem[Wiklind96]{Wiklind96}Wiklind, T., \& Combes, F. 1996, Nature, 379, 139
\bibitem[Wiklind98]{Wiklind98}Wiklind, T., \& Combes, F. 1998, ApJ, 500, 129
\bibitem[Wooden99]{wooden99} Wooden, D. H., Harker, D. E., Woodward, C. E., Butner, H. M., Koike, C., Witteborn, F. C., \& McMurtry, C. W. 1999, ApJ, 517, 1034
\bibitem[Xie14]{Xie14} Xie, Y., Hao, L., \& Li, A. 2014, ApJ, 794, L19
\bibitem[York06a]{York06a} York, D. G., Straka, L, A., Bishof, M., Kuttruff, S., Bowen, D., Kulkarni, V. P., Subbarao, M., Richards, G., Vanden Berk, D., Hall, P. B., Heckman, T., Khare, P., Quashnock, J., Ghering, L., \& Johnson, S.  2006a, MNRAS, 367, 945
\bibitem[York06b]{York06b} York, B. A., Ellison, S. L., Lawton, B., Churchill, C. W., Snow, T. P., Johnson,
R. A., \& Ryan, S. G. 2006b, ApJ, 647, L29
\bibitem[Zafar12]{Zafar12} Zafar, T., Watson, D., El\'iasd\'ottir, A., Fynbo, J. P. U., Kruhler, T., Schady, P., Leloudas, G., Jakobsson, P., Thone, C. C., Perley, D. A., Morgan, A. N., Bloom, J., \& Greiner, J. 2012, ApJ, 753, 82
\end{thebibliography}
\end{document}